\documentclass[aps,prl,10pt,twocolumn,superscriptaddress]{revtex4-2}

\usepackage{graphicx}
\usepackage{amssymb,amsmath}
\usepackage{color}

\newcommand{\bfr}{{\bf r}}

\newcommand{\bfk}{{\bf k}}
\newcommand{\bfA}{{\bf A}}

\newcommand{\bfG}{{\bf G}}

\newcommand{\bfj}{{\bf j}}
\newcommand{\bfz}{{\bf 0}}

\begin{document}

\title{Real-time exciton dynamics with time-dependent density-functional theory}

\author{Jiuyu Sun}
\affiliation{Department of Physics and Astronomy, University of Missouri, Columbia, Missouri 65211, USA}
\affiliation{Max Planck Institute for the Structure and Dynamics of Matter, 22761 Hamburg, Germany}

\author{Cheng-Wei Lee}
\affiliation{Department of Materials Science and Engineering, University of Illinois at Urbana-Champaign, Urbana, Illinois 61801, USA}

\author{Alina Kononov}
\affiliation{Department of Physics, University of Illinois at Urbana-Champaign, Urbana, Illinois 61801, USA}

\author{Andr\'e Schleife}
\affiliation{Department of Materials Science and Engineering, University of Illinois at Urbana-Champaign, Urbana, Illinois 61801, USA}
\affiliation{Materials Research Laboratory, University of Illinois at Urbana-Champaign, Urbana, Illinois 61801, USA}
\affiliation{National Center for Supercomputing Applications, University of Illinois at Urbana-Champaign, Urbana, Illinois 61801, USA}

\author{Carsten A. Ullrich}
\affiliation{Department of Physics and Astronomy, University of Missouri, Columbia, Missouri 65211, USA}

\date{\today }

\begin{abstract}
Linear-response time-dependent density-functional theory (TDDFT) can describe excitonic features in the optical spectra of insulators and semiconductors, using exchange-correlation (xc) kernels behaving as $-1/k^{2}$ to leading order. We show how excitons can be modeled in real-time TDDFT, using an xc vector potential constructed from approximate, long-range corrected xc kernels.  We demonstrate for various materials that this real-time approach is consistent with frequency-dependent linear response, gives access to femtosecond exciton dynamics following short-pulse excitations, and can be extended with some caution into the nonlinear regime.
\end{abstract}

\maketitle

{\em Introduction.}---Optical spectra of electronic systems can be calculated from first principles in two alternative ways:
Using frequency-dependent linear response (LR) theory, or via real-time (RT) propagation of the electronic wave function following a short initial excitation and
then Fourier transforming the induced current fluctuations \cite{Onida2002,Sander2017}. A RT description of the electron dynamics has
several benefits: for large systems it becomes computationally advantageous over the LR formalism \cite{Muller2020}, it allows coupling to nuclear dynamics \cite{Rozzi2013},
and it gives easy access to ultrafast (as/fs) or nonlinear processes \cite{Li2020}.

To describe the dynamics of interacting electrons, time-dependent density-functional theory (TDDFT) is an accurate yet
computationally efficient choice \cite{Runge1984,Ullrich2012,Casida2012}. Here, our interest is in the RT electron dynamics of optically
excited periodic solids with a band gap. RT-TDDFT for solids has a history of over two decades \cite{Bertsch2000,Yabana2006}.
Besides the calculation of optical spectra, it has been used to simulate two-photon absorption and ultrafast dielectric response \cite{Otobe2008,Yabana2012,Su2017,Zhang2017},
coherent phonons and stimulated Raman scattering \cite{Yamada2019}, ultrafast laser-induced metal-insulator transitions \cite{Wachter2014},
nonlinear optical response and high-order harmonic generation \cite{Goncharov2013,Tancogne2017a,Tancogne2017b}, photoelectron spectroscopy \cite{Giovannini2017},
electronic stopping power \cite{Pruneda2002,Schleife2015}, ultrafast demagnetization of ferromagnets and magnons \cite{Krieger2015,Tancogne2020},
as well as core excitations \cite{Pemmaraju2018a,Pemmaraju2018b,Pemmaraju2020}.

In Refs. \cite{Bertsch2000,Yabana2006,Otobe2008,Yabana2012,Su2017,Zhang2017,Yamada2019,Wachter2014,Goncharov2013,Tancogne2017a,Tancogne2017b,Giovannini2017,Pruneda2002,Schleife2015,Krieger2015,Tancogne2020,Pemmaraju2018a},
(semi)local exchange-correlation (xc) functionals were used, i.e., the adiabatic local-density
approximation (ALDA) or generalized gradient approximations (GGA). This causes a serious problem for semiconductors and insulators: (semi)local xc approximations cannot describe excitons \cite{Onida2002,Ullrich2015}, and therefore produce physically wrong optical absorption spectra.
Excitonic features can be captured in RT using hybrid functionals \cite{Sander2017,Pemmaraju2018b,Pemmaraju2020} or the Bethe-Salpeter equation (BSE) \cite{Attaccalite2011}.
However, these methods are computationally much more demanding than pure xc density functionals.

In this paper, we develop an RT-TDDFT approach that is capable of describing excitonic effects.
The idea is to generalize the so-called long-range corrected (LRC) xc kernels from LR-TDDFT \cite{Reining2002,Botti2004,Byun2017b,Byun2020} into the RT regime;
the result is an xc vector potential that accounts for the long-range screened electron-hole interaction that causes the formation of excitons.
We implement this approach in the Qb@ll code \cite{qball,Schleife2014,Draeger2017}, and demonstrate that it produces optical spectra that are consistent
with those obtained via LR. We then present several applications that illustrate the capabilities and limitations of this approach, including ultrafast and nonlinear effects.

{\em Theoretical background.}---In the frequency-dependent LR-TDDFT formalism, interacting electronic systems respond
to the sum of external perturbation plus linearized Hartree and xc potentials. The latter are determined by the Hartree kernel $f_{\rm H}(\bfr,\bfr') = 1/|\bfr-\bfr'|$
and the xc kernel $f_{\rm xc}(\bfr,\bfr',\omega)$; the xc kernel---a functional of the ground-state density $n_{\rm gs}(\bfr)$---has to be approximated in practice.
The LR-TDDFT formalism is widely used for calculating excitation energies and optical spectra \cite{Ullrich2012,Casida2012}.

In a periodic solid, optical absorption is defined with respect to the total macroscopic classical perturbation acting on it,
including the macroscopic classical induced field \cite{Onida2002,Reining2016}. LR-TDDFT accounts for this via a modified Hartree kernel, which
in reciprocal space is given by \cite{Byun2020}
\begin{equation}\label{fH}
f^{\rm mod}_{{\rm H}, \bfG\bfG'}(\bfk) = \frac{4\pi }{|\bfk + \bfG|^2}\,\delta_{\bfG,\bfG'}(1 - \delta_{\bfG, \bfz}) \:.
\end{equation}
Here, $\bfG,\bfG'$ are reciprocal lattice vectors, and $\bfk$ is a wavevector in the first Brillouin zone. The modification thus consists in setting the head
of the Hartree kernel (the term with $\bfG=\bfG'=0$)  to zero.

Since we are interested in optical excitations, the xc kernel $f_{{\rm xc}, \bfG \bfG'}(\bfk,\omega)$ is needed in the limit $\bfk\to 0$.
It is a known analytic property that in this limit the head of the xc kernel diverges as $k^{-2}$, the wing elements ($\bfG=0$, $\bfG'$ finite and vice versa)
diverge as $k^{-1}$, and the body elements ($\bfG,\bfG'$ finite) approach a constant \cite{Ghosez1997,Kim2002}. In three-dimensional bulk solids, the $k^{-2}$ behavior of the head of $f_{{\rm xc}, \bfG \bfG'}(\bfk,\omega)$ is the dominant effect causing the formation of excitons in LR-TDDFT \cite{Onida2002}.
Several approximations which capture this behavior have been proposed in the literature \cite{Byun2017b,Byun2020}, most of them independent of $\omega$
(adiabatic approximation). Here, we consider the simplest of these, the LRC xc kernel \cite{Reining2002,Botti2004}:
\begin{equation}\label{fxc_LRC}
f^{\rm LRC}_{{\rm xc}, \bfG\bfG'}(\bfk) = -\frac{\alpha }{|\bfk + \bfG|^2}\,\delta_{\bfG,\bfG'}\:,
\end{equation}
where $\alpha$ is, in principle, a functional of $n_{\rm gs}$, but here we treat it as a material-dependent empirical parameter.
With a suitable choice of $\alpha$, the LRC kernel can reproduce the main excitonic features
in the optical absorption spectra of insulators and semiconductors, including strongly bound and continuum excitons \cite{Byun2020}.
In the following, we limit ourselves
to the head-only LRC kernel, i.e., we set $f^{\rm LRC}_{{\rm xc}, \bfG\bfG'}(\bfk) = 0$ unless $\bfG=\bfG'=\bfz$.

The ALDA lacks the long-range ($\bfk\to 0$) behavior that is required for an excitonic xc kernel; however, it does contribute short-range
local-field effects, which can have an impact on the spectral shape. We will take advantage of this by
defining a combined xc kernel as follows:
\begin{equation}\label{dual_fxc}
f^{\rm LRC_+}_{\rm xc} =  f^{\rm LRC}_{{\rm xc}}+ \beta f_{\rm xc}^{\rm ALDA} ,
\end{equation}
where $\beta$ is an adjustable parameter which gives us some flexibility to improve LRC spectral features, if needed.

Formally, the xc kernel in LR-TDDFT is defined as the functional derivative of the time-dependent xc potential $v_{\rm xc}(\bfr,t)$. In the case of the ALDA,
this becomes $f_{\rm xc}^{\rm ALDA}(\bfr,\bfr') = \left. \delta v_{\rm xc}^{\rm LDA}[n](\bfr)/\delta n(\bfr')\right|_{n_{\rm gs}(\bfr)}$.
However, for excitonic xc kernels such as the so-called bootstrap kernel \cite{Sharma2011}, no comparable relation exists. It is a common characteristic of
most excitonic xc kernels currently in use \cite{Byun2017b,Byun2020} that they are not defined as the functional derivative of an xc potential.
It is thus not immediately obvious how to go from LR- to RT-TDDFT for this class of functionals; however, for the simple LRC xc kernel (\ref{fxc_LRC})
it is relatively straightforward, as we shall now discuss.

Consider the general situation where a solid is initially in the ground state associated with a periodic lattice potential $v(\bfr)$. We assume that the band structure
has been calculated using the LDA or any of the standard semilocal approximations (which may underestimate the band gap, but this is not a major concern here).
At time $t=0$, a time-dependent perturbation is switched on, in the form of a scalar potential $v'(\bfr,t)$
and/or a vector potential $\bfA'(\bfr,t)$. Formally, this requires the framework of time-dependent current-DFT \cite{Ullrich2012}, featuring
time-dependent xc scalar and vector potentials $v_{\rm xc}(\bfr,t)$ and $\bfA_{\rm xc}(\bfr,t)$,
and the system evolves under the time-dependent Kohn-Sham equation in the velocity gauge:
\begin{eqnarray}\label{KSA}
\lefteqn{
i \frac{\partial}{\partial t}\varphi_j(\bfr,t) = \bigg[\frac{1}{2}\left(\frac{\nabla}{i} + \bfA'(\bfr,t)+ \bfA_{\rm xc}(\bfr,t) \right)^2} \nonumber\\
&&{}  +  v(\bfr) + v'(\bfr,t) + v_{\rm H}(\bfr,t) + v_{\rm xc}(\bfr,t) \bigg]\varphi_j(\bfr,t) \:.
\end{eqnarray}

The time-dependent density can be written as
\begin{equation}\label{tdd}
n(\bfr,t) = n_{\rm gs}(\bfr) + \delta n(\bfr,t)\:,
\end{equation}
where the density response $\delta n(\bfr,t)$ is not necessarily small compared to the lattice-periodic $n_{\rm gs}(\bfr)$.
Recalling that the optical response requires removing the long-range ($\bfG=0$) part of the classical Coulomb interaction,
the time-dependent Hartree potential takes the form
\begin{equation}
v_{\rm H}(\bfr,t) = v_{\rm H}[n_{\rm gs}](\bfr) + v_{\rm H}^{\rm mod}[\delta n](\bfr,t)\:,
\end{equation}
using the modified Hartree kernel of Eq. (\ref{fH}).

Next, we consider the time-dependent xc effects. The ALDA xc potential $v_{\rm xc}^{\rm ALDA}[n](\bfr,t)$ matches the ground-state LDA,
but does not give rise to excitonic binding. To generate excitons we include additional, purely dynamical xc effects based on
the LRC kernel (\ref{fxc_LRC});  this immediately results in an LRC xc scalar potential
of the real-space form $v_{\rm xc}^{\rm LRC}(\bfr,t) = \int d\bfr' f_{\rm xc}^{\rm LRC}(\bfr,\bfr')\delta n(\bfr',t)$ \cite{Williams2021}.
The reciprocal-space form of this is
\begin{equation}
v_{{\rm xc},\bfG}^{\rm LRC}(t) = -\frac{\alpha}{|\bfG|^2}\, \delta n_\bfG(t) \:,
\end{equation}
making use of the lattice periodicity of the density response. However, the long-range $(\bfG=0)$ component
of $v_{{\rm xc},\bfG}^{\rm LRC}(t)$ is ill-defined, in spite of the fact that $\delta  n_{\bfz}(t)=0$ due to charge conservation.
This problem can be avoided by transforming into an xc vector potential \cite{Maitra2003}. In real space, we obtain
\begin{equation}\label{AXC_full}
\bfA_{\rm xc}^{\rm LRC}(\bfr,t) = -\frac{\alpha}{4\pi} \int_0^t dt' \int_0^{t'} dt'' \nabla\! \int d\bfr' \frac{\nabla' \cdot \bfj(\bfr',t'')}
{|\bfr- \bfr'|} \:,
\end{equation}
where the current density $\bfj(\bfr,t)$ enters via the continuity equation
$\nabla\cdot \bfj(\bfr,t) = -\partial n(\bfr,t) /\partial t$, and the scalar and vector potentials are connected through the gauge relation
$\partial \bfA_{\rm xc}^{\rm LRC}(\bfr,t)/\partial t = -\nabla v_{\rm xc}^{\rm LRC}(\bfr,t)$.

Since the head of the LRC xc kernel (\ref{fxc_LRC}) is dominant in the calculations of optical excitations,
we only include the macroscopic current density $\bfj_\bfz$ in the LRC vector potential \footnote{This is different in lower dimensions, where the head of the LRC
kernel is ineffective and excitons are formed via local-field effect. Hence, in Ref. \cite{Williams2021} the xc scalar potential is used}.
Thus, we end up with the following reciprocal-space form of the LRC vector potential:
\begin{equation}\label{vp1}
\bfA^{\rm LRC}_{{\rm xc},\bfG}(t) = \alpha \int_0^t dt' \int_0^{t'} dt''\bfj_{\bfG}(t'') \delta_{\bfG,\bfz}\:,
\end{equation}
which can also be written as a differential equation:
\begin{equation}\label{vp2}
\frac{d^2}{dt^2} \bfA^{\rm LRC}_{{\rm xc},\bfz}(t) = \alpha \bfj_{\bfz}(t) \:.
\end{equation}
The total current density is the sum of the paramagnetic current density $\bfj_p = (2i)^{-1}\sum_j\varphi_j^*(\bfr,t)\nabla \varphi_j(\bfr,t) + c.c.$
and a diamagnetic contribution featuring the vector potentials. Thus, the macroscopic total current density is
\begin{equation}\label{mac_j}
\bfj_\bfz(t) = \bfj_{p,\bfz}(t) + (\bfA'_{\bfz}(t) + \bfA_{\rm xc,\bfz}^{\rm LRC}(t))n_{\rm gs,\bfz}\:,
\end{equation}
where the average ground-state density $n_{\rm gs,\bfz}=N/V_{\rm cell}$ is the number of electrons per unit cell divided by the unit cell volume.

In Eq. (\ref{dual_fxc}) we introduced the LRC$_+$ kernel, combining the LRC xc kernel with the $\beta$-scaled ALDA xc kernel.
An RT-TDDFT description of excitonic effects that is consistent with this is
achieved by using the LRC xc vector potential $\bfA_{\rm xc}^{\rm LRC}(t)$ and a $\beta$-scaled scalar ALDA xc potential,
$v_{\rm xc,\beta}^{\rm ALDA}(t)$, in the Kohn-Sham equation (\ref{KSA}). The $\beta$-scaling only affects the response part of $v_{\rm xc}^{\rm ALDA}(t)$
associated with $\delta n(t)$; see Supplemental Material (SM) \cite{supp} for more details.
As before, we shall refer to this combined RT-TDDFT approach as LRC$_+$.


{\em Results and discussion.}---In the following, we present results for Si, LiF, CsGeCl$_3$, and an H$_2$ chain. The RT-TDDFT calculations
were done with Qb@ll \cite{qball,Schleife2014,Draeger2017}, and we compare with LR-TDDFT and BSE calculations using Yambo \cite{Yambo2019-short} and Quantum Espresso \cite{QE-2017-short}. Computational details are given in the SM \cite{supp}.

We begin with Si, to verify the consistency between RT- and LR-TDDFT. The LRC kernel (\ref{fxc_LRC}) was originally proposed to reproduce the optical spectrum of
Si using $\alpha=0.2$ \cite{Reining2002}. Thus, we compare $f_{\rm xc}^{\rm ALDA}$ and $f_{\rm xc}^{\rm LRC_+}$ (with $\beta=1$) in LR.
We also solve Eq. (\ref{KSA}) using the corresponding ALDA and LRC$_+$ in a cubic cell containing 8 Si atoms. Starting from the Kohn-Sham ground state, the system is excited by a delta-peaked uniform electric field along the $z$-direction, which leads to a constant $\bfA'$ switched on at $t=0$.
The dielectric function $\varepsilon(\omega)$ is obtained from the induced current fluctuations, following Yabana {\em et al.} \cite{Yabana2006,Yabana2012}.
In order to save computational resources, we used a shifted $8\times8\times8$ Monkhorst-Pack $\bfk$-point mesh, which we have
carefully tested and found to be sufficiently accurate for Si (see SM \cite{supp}).

\begin{figure}
	\centering
\includegraphics[width=\linewidth]{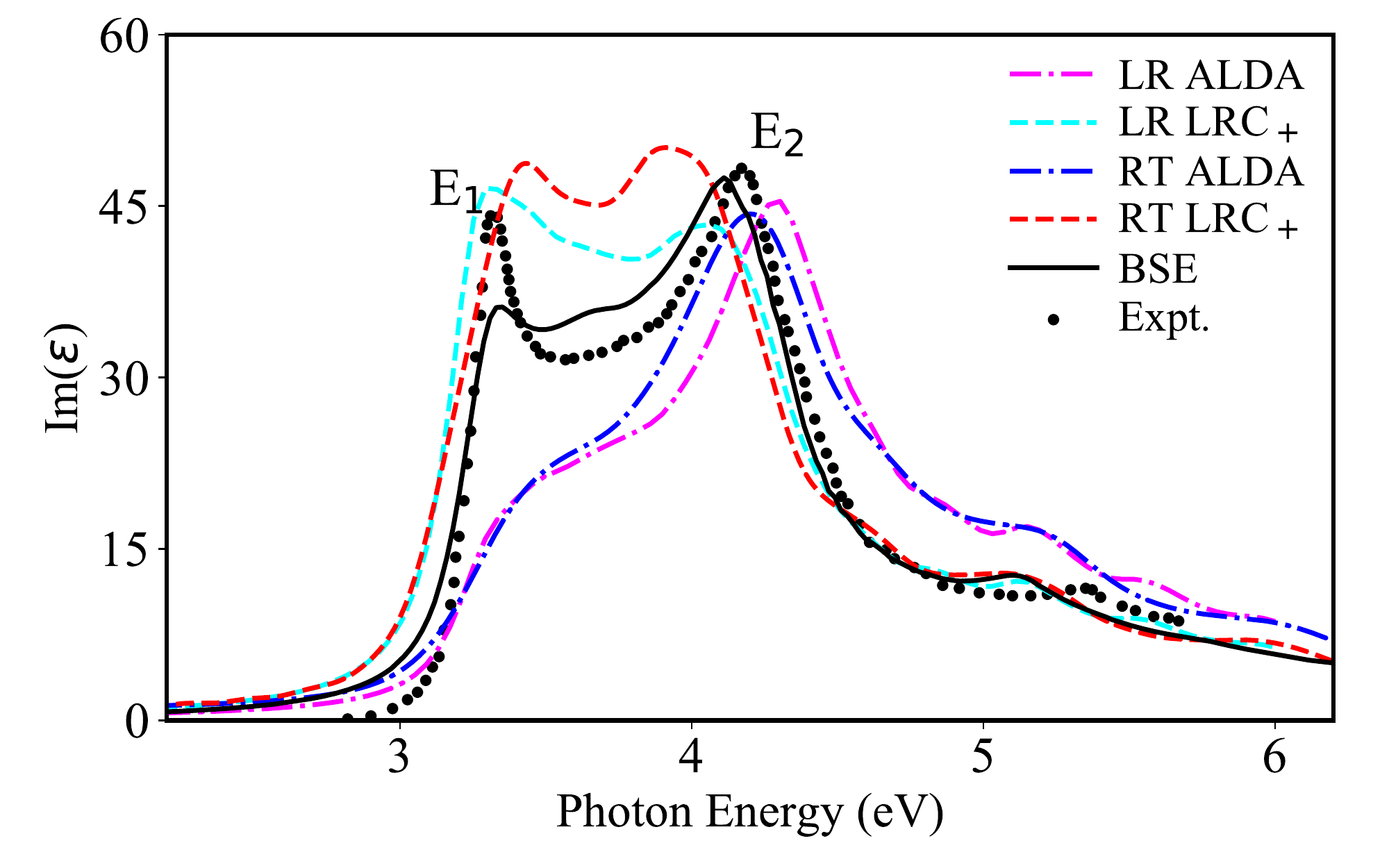}
     \caption{Optical spectra Im($\varepsilon$) of Si, obtained by LR- and RT-TDDFT, compared with BSE and experiment \cite{Lautenschlager87}.
     The calculated spectra are scissor shifted for the onset to line up with experiment (see SM for details \cite{supp}).}
     \label{fig:Si}
\end{figure}

Figure \ref{fig:Si} shows the imaginary part of the dielectric function Im($\varepsilon$) of Si obtained by different approaches, as well as experimental data. It is well known that the ALDA fails to capture the excitonic features in the optical spectrum of Si \cite{Onida2002}. As shown, the LR-ALDA and RT-ALDA spectra are very similar:
both seriously underestimate the first absorption peak $E_1$ around 3.2 eV. In BSE, the $E_1$ peak is strongly enhanced compared to ALDA, though still somewhat lower than
experiment. A better agreement between BSE and experiment could be achieved with a much denser $\bfk$-grid or other improvements \cite{Marini2003,Kammerlander2012},
but this is not the main focus of our study.

It is evident from Fig. \ref{fig:Si} that LRC dramatically improves the ALDA spectrum: both LR- and RT-LRC$_+$ curves show double-peak structures, with an $E_1$ peak height comparable to $E_2$, which agrees better with experiment than BSE. Both LRC$_+$ spectra also correct the overestimation
beyond 4.5 eV by ALDA. The differences between the LR and RT spectra are mainly due to the different $\bfk$-point sampling used in Qb@all and Yambo, as discussed in
the SM \cite{supp}. Aside from these minor technical details, our results clearly show that excitonic effects in materials with weakly bound excitons, such as Si,
can be well described with RT-TDDFT using LRC$_+$.

\begin{figure}
	\centering
\includegraphics[width=\linewidth]{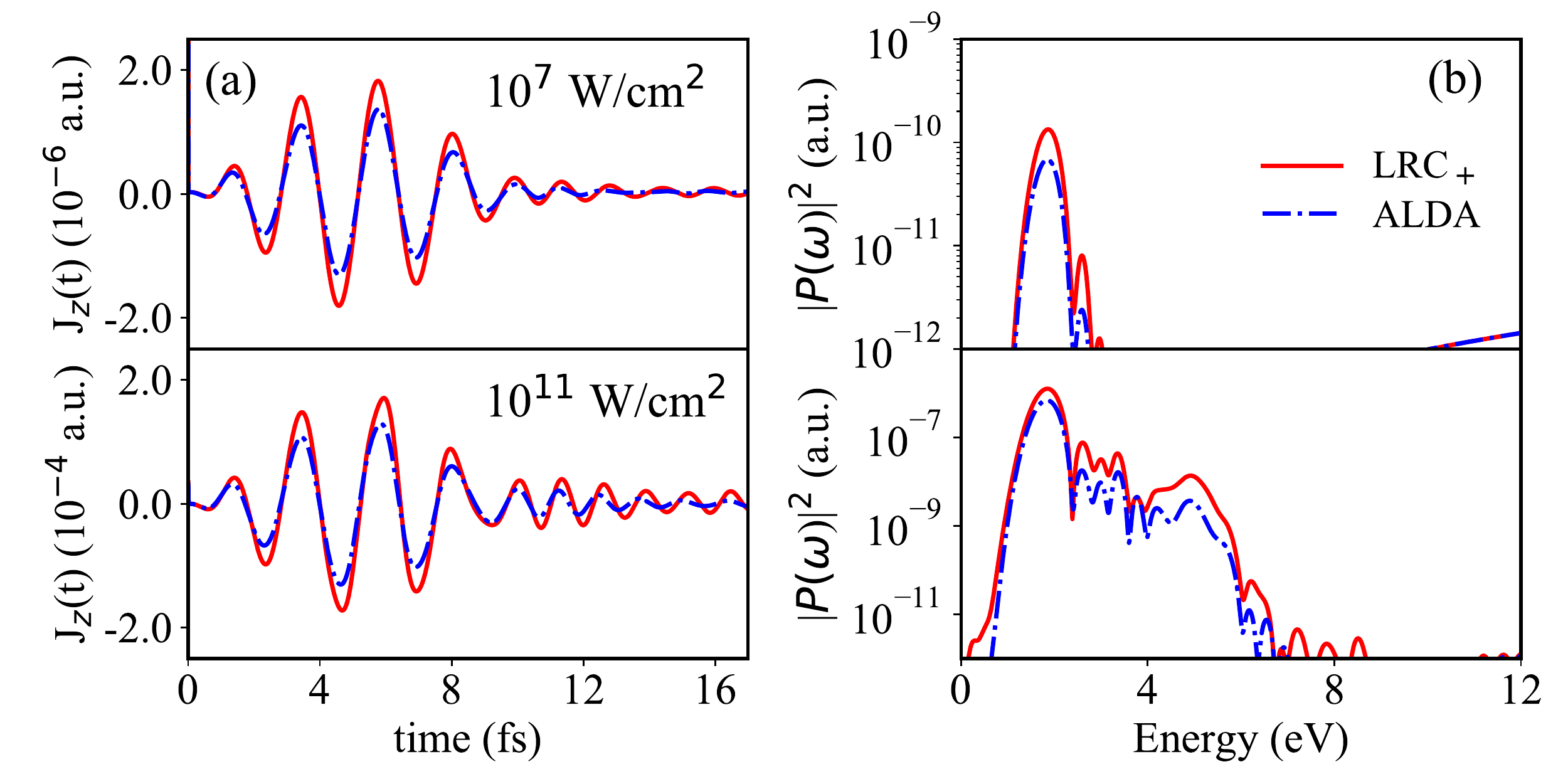}
     \caption{Response of Si to 10 fs laser pulses (frequency 1.6 eV, polarized along $z$) with peak intensities 10$^{7}$ W/cm$^2$ (top) and 10$^{11}$ W/cm$^2$
     (bottom), comparing ALDA and LRC$_+$ within RT-TDDFT. (a) Induced current density $j_z(t)$. (b) Dipole power spectrum $|P(\omega)|^2$.}
     \label{fig:laser}
\end{figure}

RT-TDDFT is not limited to weak perturbations, but allows us to explore ultrafast and nonlinear electron dynamics.
Instead of a delta-peaked uniform electric field, we apply short laser pulses polarized along the $z$-axis with a frequency of 1.6 eV, sin$^2$ envelope,
and pulse duration of 10 fs. We consider weak and strong pulses with peak intensity 10$^{7}$ and 10$^{11}$ W/cm$^2$, respectively. Figure \ref{fig:laser}a
 shows that the $z$-component of the total macroscopic current density $j_{z}^{\rm tot}$ propagated with LRC$_+$ has a larger amplitude than with ALDA.
There are two reasons for the enhanced current response: (i) LRC drastically increases the oscillator strength at the absorption edge
(see Fig. \ref{fig:Si}), leading to a stronger coupling to the laser; (ii) the diamagnetic contribution to the total current, see Eq. (\ref{mac_j}),
is enhanced by the LRC xc vector potential. While the system is driven by the laser, the induced currents scale with the square root
of the intensity; the remaining current oscillations after the end of the pulse are more pronounced at 10$^{11}$ W/cm$^2$, indicating nonlinearity.

The associated dipole power spectra $|P(\omega)|^2$ (see SM \cite{supp}) are shown in Fig.~\ref{fig:laser}b. At low laser intensity, ALDA and LRC$_+$ produce
very similar spectra, with a dominant peak at 1.6 eV and a smooth drop-off at higher frequencies. Nonlinear effects become significant at 10$^{11}$ W/cm$^2$
pulse intensity: the ALDA and LRC$_+$ spectra both extend towards higher frequencies, and there is a broad peak around 5 eV (the 3rd harmonic of the pulse).
Overall, LRC$_+$ gives a more pronounced nonlinear response than ALDA, which is in agreement with a study using time-dependent polarization-DFT \cite{Gruning2016}.

\begin{figure}
	\centering
\includegraphics[width=\linewidth]{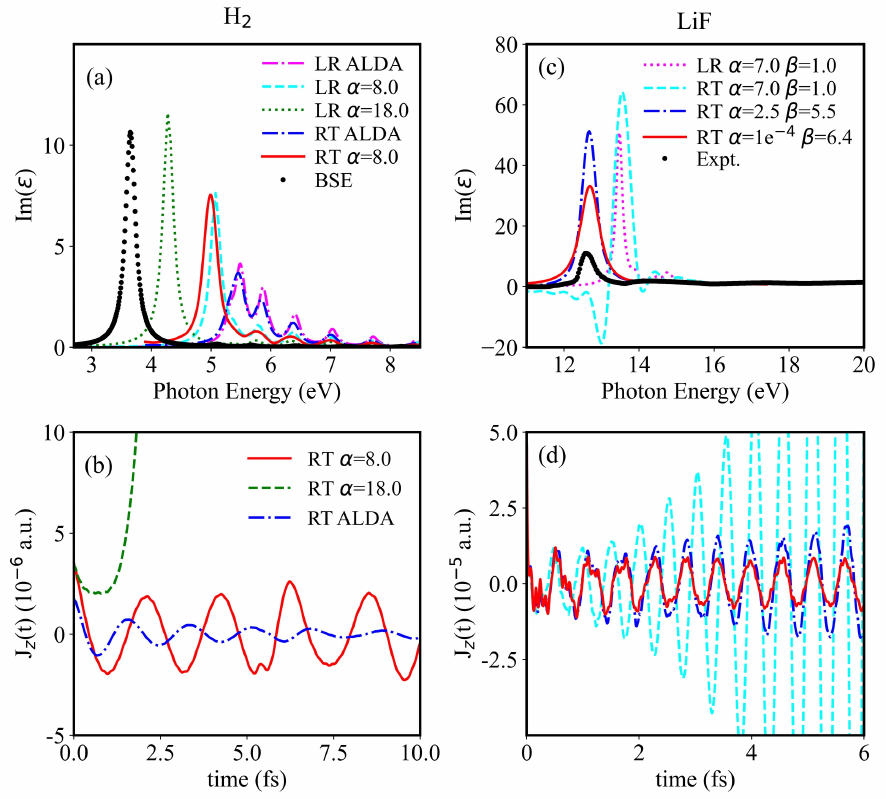}
     \caption{Strongly bound excitons in an H$_2$ chain (left) and LiF (right). (a) Im($\varepsilon$) from BSE and TDDFT; (b) macroscopic current density from
      RT ALDA and LRC with $\alpha=8$ and $18$; (c) Im($\varepsilon$) from LRC$_+$ in LR and RT, with $\alpha$ and $\beta$ as indicated, versus experiment \cite{RW67};
     (d) macroscopic current density from the same three RT-LRC$_+$ as in (c). }
     \label{fig:Bound}
\end{figure}

Next, we explore strongly bound excitons in insulators. We begin with a chain of H$_2$ molecules with a lattice constant of 4.5 a.u.
(see SM \cite{supp}). Figure \ref{fig:Bound}a shows that BSE yields an optical spectrum with a pronounced excitonic peak around 3.6 eV;
the ALDA fails to reproduce this peak. LR-TDDFT with the LRC kernel improves the spectra: for $\alpha=18.0$ we obtain an excitonic peak of similar height
and shape as the BSE, but at a higher energy. An even larger $\alpha$ would put the excitonic peak at the right position, but with too much
oscillator strength, consistent with earlier studies of the LRC kernel \cite{Byun2017b}.

For $\alpha=8.0$, LR- and RT-TDDFT of the H$_2$ chain are in close agreement. However, we found that at $\alpha=18.0$ the RT calculation
failed. Figure \ref{fig:Bound}b shows that at $\alpha=8.0$  the induced current is comparable to the ALDA current, but at $\alpha=18.0$ the
current rapidly diverges.

To investigate this further, we now consider LiF. The experimental optical spectrum (see Fig. \ref{fig:Bound}c) features a prominent excitonic peak around
12.6 eV. LR-TDDFT with LRC$_+$ using $\alpha=7$ and $\beta=1$ gives a blue-shifted exciton at 13.5 eV; a larger value of $\alpha$ could be
used to shift the exciton down to the correct position, but with much exaggerated peak height \cite{Byun2017b}.

RT-TDDFT using LRC$_+$ with the same parameters $(\alpha=7,\beta=1)$ appears to be developing an instability, as indicated by the current density in Fig. \ref{fig:Bound}d
which keeps increasing after 2 fs. The resulting LiF optical spectrum (cyan curve in Fig. \ref{fig:Bound}c) is peaked at 13.5 eV but has a distorted shape.
The current response can be stabilized by decreasing $\alpha$, and the excitonic peak can be shifted to the correct position by increasing $\beta$,
as illustrated in Figs. \ref{fig:Bound}c and d. Indeed, comparing $(\alpha=2.5,\beta=5.5)$ and $(\alpha=10^{-4},\beta=6.4)$ we find that
the latter produces the best agreement with experiment. In this case, the excitonic interactions are caused by emphasizing the local-field effects,
like in the so-called contact exciton \cite{Sottile2003,Botti2007}.

What is the reason for the LRC instabilities?
The zero-force theorem of TDDFT \cite{Ullrich2012} states that the total force due to xc scalar and vector potentials must vanish:
\begin{eqnarray}\label{zeroforce}
0 &=&
\int d\bfr\Big [ -n(\bfr,t)\nabla v_{\rm xc}(\bfr,t) - n(\bfr,t)\frac{\partial}{\partial t}\bfA_{\rm xc}(\bfr,t) \nonumber\\
&& {}+ \bfj(\bfr,t)\times\nabla\times \bfA_{\rm xc} (\bfr,t)\Big].
\end{eqnarray}
The ALDA xc potential satisfies the zero-force theorem. $\bfA^{\rm LRC}_{\rm xc}$ is strictly longitudinal, so the last term in
Eq. (\ref{zeroforce}) vanishes. From Eq. (\ref{vp1}), the second term in Eq. (\ref{zeroforce}) becomes $-\alpha N \int_0^t dt' \bfj_{\bfz}(t')$.
Thus, LRC produces a macroscopic xc force, which can cause instabilities in the current oscillations for strongly bound excitons, as we have seen in H$_2$ and LiF.
This violation of the zero-force theorem is also present in $f_{\rm xc}^{\rm LRC}$, but still allows one to obtain good optical spectra,
albeit with an exaggerated oscillator strength for strongly bound excitons \cite{Byun2017b}; the instabilities only show up in the nonlinear regime.

As a final illustration of RT-TDDFT, we now return to a system with weakly bound excitons and consider a more complex material, the perovskite CsGeCl$_3$.
To our knowledge, no experimental optical spectra of this material are available.
We adopt a cubic phase of $Pm\bar{3}m$, where a Ge atom substitutes the Pb atom in the popular CsPbCl$_3$, which allows us to neglect spin-orbit coupling.

\begin{figure}
	\centering
\includegraphics[width=\linewidth]{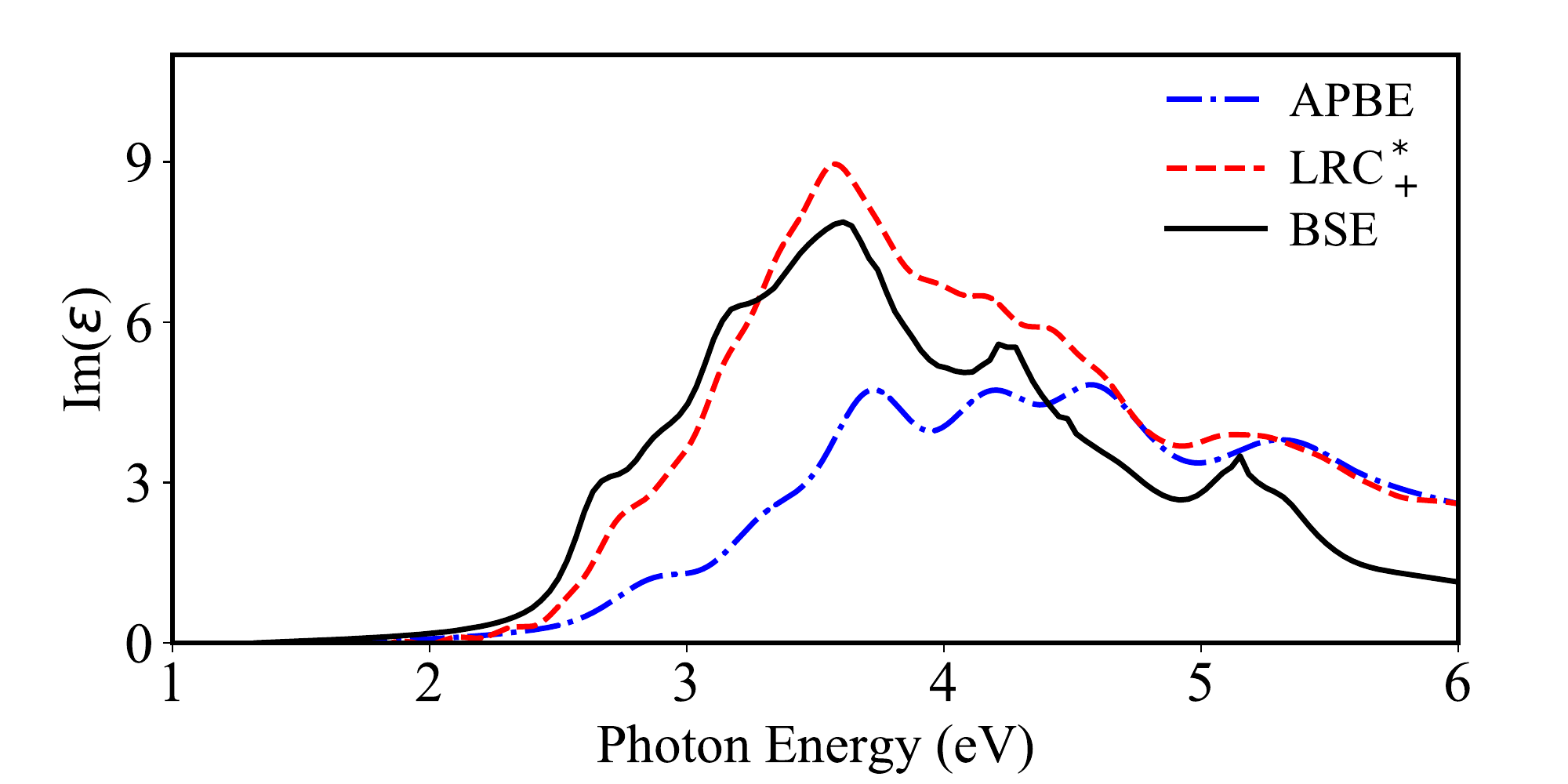}
     \caption{Optical spectra of CsGeCl$_3$ obtained by BSE and RT-TDDFT using APBE and LRC$_+^*$.}
     \label{fig:CsGeCl}
\end{figure}

Figure \ref{fig:CsGeCl} shows the optical spectrum of CsGeCl$_3$, calculated using $G_0W_0$+BSE. The $G_0W_0$ band gap is 2.96 eV;
the BSE spectrum displays a relatively weak shoulder around 2.6 eV, and a dominant continuum exciton peak at 3.5 eV.
We compare with RT-TDDFT spectra obtained using adiabatic PBE (APBE) \cite{Perdew1996} and APBE+LRC (LRC$_+^*$) using $\alpha=1.1$.

As shown in Fig. \ref{fig:CsGeCl}, the APBE and LRC$_+^*$ spectra are almost on top of each other beyond 4.6 eV, and they are both very similar to BSE in this range.
At lower energies, APBE, a semilocal functional, significantly underestimates Im($\varepsilon$); this is similar to the failure of ALDA seen for Si.
On the other hand, the overall spectral shape of LRC$_+^*$ is very close to BSE, even reproducing the weak shoulder around 2.8 eV. The associated induced current
densities (see SM \cite{supp}) are well behaved and stable during the entire time propagation.

{\em Conclusions.}---In this paper, we have demonstrated that TDDFT can describe excitons in periodic solids by propagating the time-dependent
Kohn-Sham equation following an initial short-pulse excitation. LR-TDDFT has long been known to be capable of producing excitonic optical spectra
using xc kernels with the appropriate long-range behavior. Here, we have shown how the simplest of these, the LRC kernel, can be converted into
an xc vector potential featuring the macroscopic current density and an adjustable parameter, $\alpha$.

We have applied this RT-TDDFT approach to Si, an H$_2$ chain, LiF, and CsGeCl$_3$, comparing, when appropriate, with LR-TDDFT, BSE, and experiment.
We find that LR- and RT-TDDFT are consistent, in the sense that they produce essentially the same optical spectra in the weakly perturbed regime, but
RT-TDDFT can be applied beyond the linear-response regime, to describe ultrafast and nonlinear exciton dynamics.
However, the LRC xc functional has its limitations: in materials with strongly bound excitons, it can lead to instabilities in the induced currents,
which is a consequence of violating the zero-force theorem. In materials with weakly bound or continuum excitons, no such problems occurred.

This study opens up multiple paths towards TDDFT studies of exciton dynamics in bulk materials and nanostructures. An important task will be to find xc functionals
for RT-TDDFT beyond the simple LRC approximation. Our RT-TDDFT approach can be combined with
recently developed visualization methods for exciton wave functions \cite{Williams2021}, and it is possible to study exciton relaxation effects by coupling
to nuclear dynamics at the Ehrenfest level \cite{Schleife2015}.

\acknowledgments{J.S. and C.A.U. acknowledge support by NSF grant No. DMR-1810922,
A.K. and A.S. acknowledge support by NSF Grant No.\ OAC-1740219, and C.W.L. and A.S. acknowledge support from the Office of Naval Research (Grant No.\ N00014-18-1-2605).
This work used the high-performance computing infrastructure provided by
Research Computing Support Services at the University of Missouri–Columbia, and
the Illinois Campus Cluster, operated by the Illinois Campus Cluster Program (ICCP) in conjunction with the National Center for Supercomputing Applications (NCSA),
supported the University of Illinois at Urbana-Champaign.}

\bibliography{TDLRC_paper_refs}

\end{document}